\documentclass[aps,prl,nofootinbib,floatfix,preprintnumbers,twocolumn,amsmath,amssymb]{revtex4}
\usepackage{graphicx}
\usepackage{multirow}
\usepackage[english]{babel}
\usepackage{amsmath}
\usepackage{graphicx}
\usepackage{color}
\usepackage{amssymb}
\usepackage{hyperref}
\usepackage{dcolumn}
\usepackage{bm}

\def\s{\sigma}

\def\beq{\begin{eqnarray}}
\def\eeq{\end{eqnarray}}

\newcommand{\vev}[1]{ \left\langle {#1} \right\rangle }

\newcommand{\MEV}{ {\rm MeV} }
\newcommand{\GEV}{ {\rm GeV} }
\newcommand{\TEV}{ {\rm TeV} }

\newcommand{\CM}{ {\rm cm} }
\newcommand{\SE}{ {\rm s} }

\begin{document}

\title{The right generations}

\author{
Alfredo Aranda,$^{a,b}$\footnote{Electronic address: fefo@ucol.mx}
Jose A. R. Cembranos,$^{a,b,c}$ \footnote{Electronic address: cembra@fis.ucm.es}
\vspace*{0.3cm}}

\affiliation{$^{(a)}$ Facultad de Ciencias, CUICBAS, Universidad de Colima, 28040 Colima, Mexico;}
\affiliation{$^{(b)}$ Dual C-P Institute of High Energy Physics, 28040 Colima, Mexico;}
\affiliation{$^{(c)}$ Departamento de F\'{\i}sica Te\'orica I, Universidad Complutense de Madrid, E-28040 Madrid, Spain.}

\date{\today}

\begin{abstract}
The Standard Model has three generations of fermions and although it does not contain any explicit reason for this, the existence of additional generations is now very constrained by experiment. Present measurements are saturating perturbative unitarity limits. The main idea of this work is to show that those restrictions can be relaxed if the new generations experience different interactions. This new setup leads to the presence of additional stable degrees of freedom that give rise to a very rich phenomenology for cosmology, astrophysics and particle physics. The stability is a consequence of the conservation of new accidental baryon and lepton numbers.
We present an explicit example by introducing a fourth generation charged under a
new $SU(2)_R$ gauge interaction instead of the standard $SU(2)_L$.
The simplest implementations lead to models that contain stable quarks, leptons and neutrinos.
We show that these new particles can have a wide range of masses within a non-standard cosmological
set-up. Indeed, the new neutrinos (and {\it neutral leptons}) constitute viable dark matter candidates if
they are the lightest of these new particles.
\end{abstract}
\maketitle

There have been several motivations to explore the possibility of a fourth (or more) generation(s). This has typically been done by postulating an exact (heavier) set of quarks and leptons in complete analogy with the known three generations, namely with the same {\it chiral} charges under the Standard Model (SM) gauge group. Immediate challenges to this proposal are the required heaviness of the fourth neutrino, as required by the $Z$-width, and more recently and devastating, due to the value obtained for the Higgs mass~\cite{Aad:2012tfa,Chatrchyan:2012ufa}, the difficulty in providing a large enough mass to the new quarks (basically one would need non-perturbative Yukawa couplings)~\cite{Djouadi:2012ae,Lenz:2013iha,Vysotsky:2013gfa}.  Thus, it seems that introducing a new generation has fallen out of grace.

There are also several reasons to extend the gauge structure of the SM, most of which emane from the idea of gauge coupling unification and grand unified theories (GUTs). In this regard, a particularly attractive and useful scenario is that of the so-called left-right models where an $SU(2)_R$ is added to the SM gauge group~\cite{Mohapatra:1974hk,Mohapatra:1974gc,Senjanovic:1975rk}. The basic idea is that what we observe as right handed fermions, singlets under the SM $SU(2)_L$, are really remnants of fermionic $SU(2)_R$ doublets. It just so happens that this new symmetry was broken by the vacuum expectation value (vev) of a bi-doublet in such a way that only the SM gauge group survives and its  matter content remains massless, including now an $SU(2)_L$ doublet scalar. This general picture is not only nice in terms of {\it restoring} the left-right symmetry lost in the SM, but is also easily embedded in larger grand unified models with a single big gauge group.

In this short letter we forget about all of that and present a couple of simple models where a new {\it right generation} is included and the SM gauge group is extended with an extra $SU(2)_R$ but with no regard, nor worry, about its possible implementation into a GUT. The idea is to consider the following gauge group: $SU(3)_C \times SU(2)_L \times SU(2)_R \times U(1)_X$,
where $X$ may or not denote Hypercharge (Y)~\footnote{In this setup we do not explore the possibility of gauged Baryon (B) and/or Lepton (L) numbers (nor B-L), and they are just accidental global symmetries of the Lagrangian.}. Let's first suppose it does not. We want to generate the following symmetry breaking pattern:
$SU(3)_C \times SU(2)_L \times SU(2)_R \times  U(1)_X \stackrel{ \langle H_R \rangle} {\longrightarrow}
SU(3)_C \times SU(2)_L \times  U(1)_Y\stackrel{ \langle H_L \rangle} {\longrightarrow} SU(3)_C \times  U(1)_{em}$.

We can accomplish it by introducing two scalar fields $H_R \sim (1, 1, 2, 1/2) $ and $H_L \sim (1, 2, 1, 1/2)$, where the numbers in parenthesis correspond to their charges under $SU(3)_C \times SU(2)_L \times SU(2)_R \times U(1)_X$. The idea is that the vev of $H_R$ gives the first breaking and that of $H_L$ the second.
Note that the electric charge is given by $Q = \tau_{3L} + Y = \tau_{3L} + \tau_{3R} + X$. The broken gauge boson spectrum consists of six massive gauge bosons denoted by $W_R^{\pm}$, $Z_R$ and the usual $W^{\pm}$ and $Z^{0}\equiv Z$. The mass scale of the right-gauge bosons is that of $\langle H_R\rangle$.

As for matter fields, the content is that of the SM (all SM fields being singlets under $SU(2)_R$) and a new (or more) {\it right generation(s)} (fully singlet under $SU(2)_L$) charged, in a mirror way, under $SU(2)_R$. Namely for leptons we have
\begin{eqnarray}
\label{leptons-charges}
\begin{array}{cccccc}
L_i&\sim &(1,2,1,-1/2)\,,  & R^{\prime}&\sim &(1,1,2,-1/2)\,, \\
E_{Ri}&\sim &(1,1,1,-1)\,, & E^{\prime}_{L}&\sim &(1,1,1,-1)\,,
\end{array}
\end{eqnarray}
and for quarks
\begin{eqnarray}
\label{quarks-charges}
\begin{array}{cccccc}
Q_{Li} &\sim& (3,2,1,1/6)\,, & Q^{\prime}_{R} &\sim& (3,1,2,1/6)\,, \\
U_{Ri}&\sim &(3,1,1,2/3)\,, & U^{\prime}_{L}&\sim &(3,1,1,2/3)\,, \\
D_{Ri}&\sim &(3,1,1,-1/3)\,, & D^{\prime}_{L}&\sim &(3,1,1,-1/3)\,, \\
\end{array}
\end{eqnarray}
where unprimed (primed) letters denote SM (new) fields.
The electric charge of the new fermions mirrors that of the SM fields. So far this setup has four massless fermions: the three left-handed neutrinos of the SM plus the new right-handed one present in the $SU(2)_R$ doublet $R^{\prime}$. Whatever mechanism is introduced to give mass to the SM neutrinos, it either must also give mass to the new one or else be extended to do so in the appropriate manner. The interesting point however is that the $Z$-width constraint on its mass does not apply in this case and it can be light.

At low energies, communication between the two sectors occurs through gluons, photons and the possible scalar mixing, which however can be very small in the decoupling limit $\langle H_R\rangle \gg \langle H_L\rangle$. The scalar potential for this scenario is
\begin{eqnarray}
\label{potential-charged}
V &=& \mu_L^{2}H_L^{\dagger}H_L + \mu_R^{2}H_R^{\dagger}H_R + \lambda_L \left( H_L^{\dagger}H_L \right)^{2} + \nonumber \\
& + & \lambda_R \left( H_R^{\dagger}H_R \right)^{2} + \lambda_{LR }\left( H_L^{\dagger}H_L \right)  \left( H_R^{\dagger}H_R \right) \ .
\end{eqnarray}

Let us now consider the second possibility alluded above, namely that $X$ is already Hypercharge: $X=Y$. In this case the breaking proceeds as $SU(3)_C \times SU(2)_L  \times  SU(2)_R \times   U(1)_Y \stackrel{ \langle H_R \rangle} {\longrightarrow} SU(3)_C \times SU(2)_L \times  U(1)_Y \stackrel{ \langle H_L \rangle} {\longrightarrow}  SU(3)_C \times  U(1)_{em}$, where this time  $H_R \sim (1, 1, 2, 0) $ and $H_L \sim (1, 2, 1, 1/2) $. Note that in this case the electric charge is given by $Q = \tau_{3L} + Y$. Anomaly cancelation, together with the fact that $H_R$ has $Y=0$, make the Hypercharge of the new leptons proportional to the one of the new quarks. Thus there are two general scenarios in this case: either all new fermions are electrically charged or all are neutral.
The scalar potential is the same as before and, for the case of a neutral generation, the only {\it direct} connections at low energies between the two worlds is through $SU(3)_C$ and $\lambda_{LR}$.

The phenomenology of these {\it right generations} is very different from the standard fourth generation models. Indeed, in the minimal framework, if the new gauge bosons are heavier than the new fermions, the latter are stable due to the conservation of the right baryon and lepton numbers. In principle, this fact makes it difficult to think that these models are viable for relatively low mass scales due to cosmological constraints
since they contain stable quarks and charged leptons. Indeed, the new quarks $q^{\prime}$ will be confined with each other and with ordinary charged quarks $q$ into color-singlet states of new mesons ([$\bar q' q$], [$q'\bar q$], and [$\bar q' q'$]) and into
baryons ([$q'qq$], [$q'q'q$], and [$q'q'q'$] depending on the number of new quarks). The lightest flavor of each of these new states can be stable
(as for example, [$\bar q' u$] and [$q'ud$]). These resulting hadrons can be electrically charged or neutral. For example, if the new generation is neutral, all the new hadrons containing standard quarks will have associated fractional electric charges. On the contrary, the new leptons will be electrically charged except
for the neutral generation. In all these cases, the associated relic abundances must be very suppressed. In general, new hadronic states will form heavy nuclei. A positively charged particle $X^+$ can capture an electron to form a bound state
chemically similar to a heavy hydrogen atom. An $X^-$ can instead bind to an
$\alpha^{++}$ particle and an electron, resulting again in a heavy hydrogen-like atom
\footnote{
Alternatively, it can capture a proton to produce a bound state.}.
A fractionally charged particle can also form {\it heavy hydrogen atoms}
by grouping a larger number of these new particles \footnote{On Earth they can also bind to more nuclei to form other anomalous isotopes of heavier nuclear species. Indeed, searches for these isotopes are more suitable to constraint neutral and stable Strongly Interacting Massive Particles (SIMPs).
However null-searches for other nuclei species than hydrogen are less constraining  \cite{PDG, Hammick}.
We will focus on the strongest bounds to stable $+1$ CHArged Massive Particles (CHAMPs) in order
to simplify the discussion and to clearly show the allowed wide range of masses accesible to the new stable
particles.}.

In these cases, most stringent bounds on their abundances come from searches of anomalous heavy water, since these new particles will form states chemically equivalent to hydrogen but with a larger mass. The relative density of these particles $n_X$, trapped in oceans and lakes with respect to standard hydrogen $n_H$, can be estimated from their cosmological abundance $\Omega_X$ as \cite{Kudo Yamaguchi}:
\begin{eqnarray}
\left(
\frac{n_X}{n_H}\right)_{Earth} \simeq c_X \cdot 10^{-5}
\left(\frac{\mbox{1\,\GEV}}{M_X}\right)\Omega_{X}h^2\,,
\end{eqnarray}
where $h$ is the dimensionless Hubble parameter and  $c_X$ is an order of one number that takes into account possible different assumptions ($c_X\sim 4$ if the new particles are
heavy $M_X\gtrsim 20\, \TEV$, and remain suspended in the Galactic halo; whereas $c_X\sim 7$ if they are light $M_X\lesssim 20\, \TEV$,
and mainly present in the Galactic disk \cite{Kudo Yamaguchi}).
The non-observation of these new heavy particles in searches of anomalous hydrogen constrains their relative density to be very diluted
\footnote{Eq. (\ref{constraints}) gives a very good approximation to the strongest constraints coming from
concentration of heavy stable particles in matter with charge $+1$. Read \cite{PDG} and references within for a comprehensive discussion of the constraints on the mass and charge of the new particle.}
\cite{PDG}:
\begin{eqnarray}
\label{constraints}
\left(\frac{n_X}{n_H}\right)_{Earth}
\left\{
\begin{array}{c}
  < 2\cdot 10^{-28} \,,\;\;       10\, \GEV \leq M_X \leq 1\, \TEV\,; \\
  < 1\cdot 10^{-14} \,,\;\;\;  1\,  \TEV  <   M_X\,. \;\;\;\;\;\;\;\;\;\;\;\;\;\;\;\;
\end{array}
\right.
\end{eqnarray}

However, these restrictions can be avoided if low reheat or maximum temperatures with respect to the new particle masses are considered within the inflationary framework. If we assume that the production is dominated by scattering processes in the thermal bath and not by direct
inflaton decays, the abundances associated with these new states can be efficiently suppressed.
In order to compute the number density $n$, of any of the new stable particles, we can use the Boltzmann equation:
\begin{eqnarray}
\label{Boltzmann}
\frac{d}{dt} n + 3 H n=-  \vev{\s v}(n^2 - n_{\rm EQ}^2) \ ,
\end{eqnarray}
where $\vev{\s v}$ is the thermal averaged annihilation cross section times velocity, $H$ is the Hubble parameter, and
$n_{\rm EQ}$ is the corresponding thermal equilibrium number density.
Neglecting a possible back-reaction from the thermal bath, the Hubble parameter and the temperature of the universe $T$
are determined by the inflaton energy density, so the dependence on the scale factor $a$, after the end of inflation
is \cite{Chung:1998zb,Chung:1998ua,Chung:1998rq,Giudice:2000ex,Feldstein:2013uha}:
\begin{eqnarray}
\left(\frac{H}{H_R}\right)^2
= \left(\frac{T}{T_R}\right)^2
= \left(\frac{a}{a_R} \right)^{-3}\,,
\end{eqnarray}
where the subscript $R$ means the value at the end of the reheating stage.
By assuming that these particles have not thermalized at any time ($n\ll n_{\rm EQ}$),
we can estimate their present abundance as:
\begin{eqnarray}
\Omega_{0}h^2 \simeq
\frac{s_0 g^2 x_R^{-7} }{36 \pi^6 H_0^2 M_{\rm pl} }
\left( \frac{90}{g_*}\right)^{3\over 2}
\mathcal{F}(x_{\rm max})\,,
 \label{abundanceNT}
\end{eqnarray}
where $M_{\rm pl}\equiv (8 \pi G_N)^{-1/2}\simeq 2.4 \cdot 10^{18}\, \GEV$ is the reduced Planck scale,
$g$ the number of degrees of freedom associated with the stable particle (for instance, $g=12$ for a quark),
$g_*$ the effective number of relativistic degrees of freedom
produced by the reheating mechanism (for example, $g_*=106.75$ accounts for all the SM particles,
which are relativistic for $T\gtrsim 300\, \GEV$),
"${\rm max}$" denotes the maximum temperature reached by the thermal bath,
$x=M/T$ (where $M$ denotes the mass of the stable particle),
$H_0 = 100$\,km/s Mpc$^{-1}$,
$s_0\simeq 2890\, \text{cm}^{-3}$ the present entropy density, and
\begin{eqnarray}
\mathcal{F}(y)=
M^2\int_{y}^\infty
\vev{\s v} x^8 e^{-2 x}\, dx\,.
\end{eqnarray}
Typically, the annihilation cross section is dominated by a particular wave channel characterized by
an integer number $j$ as:
\begin{eqnarray}
\vev{\s v}\simeq M^{-2}c_j x^{-j}\,,
\end{eqnarray}
so $\mathcal{F}(y)$ can be written in terms of the incomplete Gamma function:
\begin{eqnarray}
\mathcal{F}(y)\simeq \frac{\Gamma(9-j,2y)}{2^{9-j}}c_j
\simeq\left\{
\begin{array}{c}
  \frac{(8-j)!}{2^{9-j}}c_j\,,\;\; y \ll 3\,; \\
  \frac{y^{8-j}}{2 e^{2 y}} c_j\,,\;\; y \gg 3\,.
\end{array}
\right.
\end{eqnarray}
Therefore, if $M \lesssim 3\, T_{\rm max}$, the abundance is suppressed by $(T_{R}/M)^7$
(by assuming $M \gg T_R$) due to an important entropy production before reheating.
On the other hand, if $M\gg T_{\rm max}$, the abundance is exponentially suppressed
by the Boltzmann statistical factor. In this latter case, the heavy water constraints on
the masses of the new stable particles are weaker. 
The s-wave ($j=0$) quark annihilating channel into gluons gives:
\begin{eqnarray}
c_0^{\;q}\simeq
\frac{7 \pi}{54}
\alpha_s^2\,,
\label{cQ}
\end{eqnarray}
where the exact value of the strong fine-structure constant $\alpha_s$, depends on $T$.
By taking into account Eq (\ref{cQ}) and by assuming $T_{\rm max}\simeq T_R$,
we can estimate the aforementioned restrictions as:
\begin{eqnarray}\label{MvsT}
M_X \gtrsim k_X\,T_{\rm max} \ln\left(\frac{T_{\rm max}}{1\,\GEV}\right)\,,
\end{eqnarray}
where $k_X\simeq 53$ for $10\, \GEV \leq M_X \leq 1\, \TEV$ whereas $k_X\simeq 37$ for $M_X > 1\;\TEV $
\footnote{Eq.~\eqref{MvsT} estimates the constraints to stable quarks assuming their mass and that of the hadrons they form to be the same. The bounds on
electrically charged particles are similar, but slightly less constraining in general, due to their
weaker coupling.}.
Concrete restrictions about the maximum temperature have not been established, but
its value is bounded from below by the reheat temperature. The agreement of the observations of primordial
abundances with the predictions of the Big Bang nucleosynthesis model implies $T_R\gtrsim 2\, \MEV$ \cite{Kawasaki:1999na,Ichikawa:2005vw}.
This constraint can be slightly improved by taking into account more cosmological data ($T_R\gtrsim 4\, \MEV$ \cite{Hannestad:2004px}).
In any case, the above estimation shows that present cosmological constraints are not competitive with respect to laboratory or
high energy experiments since they can be fulfilled for the entire range of masses where they are defined ($M_X \gtrsim 10\, \GEV$). It is important to note that the restrictions are weaker in our analysis with respect to other works (\cite{Kudo Yamaguchi}, for instance), due to our assumption of a low maximum temperature after inflation, and not only a low reheat temperature.

On the other hand, the new neutrino, or the lepton if neutral, can have associated important cosmological
abundances without conflicting present observations. Indeed, they can be the main component of the
non-baryonic dark matter (DM). For example, for masses much higher than the electroweak phase transition, 
we can compute the annihilation cross section into the the total
bosonic and fermionic content of the SM due to Hypercharge interactions. For the case of a Dirac fermion:
\begin{eqnarray}
\label{eq:ann}
c_0^{\;Y}&\simeq& \frac{\pi}{8}
Y^2 (41 + 8 Y^2)\,\alpha_Y^2\,,
\label{c0Y}
\end{eqnarray}
which is again s-wave dominated. $Y$ is the Hypercharge of the new stable particle.
It is expected to be order one, and for concreteness, we will assume $Y=1/2$. In such
a case $c_0^{\;Y}\simeq 4.4\cdot 10^{-4}$.
Following the standard
{\it freeze-out} computation \cite{Srednicki:1988ce,Cembranos:2003mr,Cembranos:2003fu},
we obtain the following semi-analytical approximated solution (for $M \gtrsim 100\,\GEV$)
for Eq. (\ref{Boltzmann}):
\begin{eqnarray}
\Omega_{0}h^2 &\simeq& 0.12\,
\left(
\frac{g}{4}
\right)
\left(
\frac{5.0\,\cdot 10^{-26} \CM^3 \SE^{-1}}{\vev{\s v}_f}
\right)
\nonumber\\
& &
\left[1-0.10\,\ln\left(
\frac{5.0\,\cdot 10^{-26} \CM^3 \SE^{-1}}{\vev{\s v}_f}
\right)\right]\,,
 \label{abundanceT}
\end{eqnarray}
where $\vev{\s v}_f$ is $\vev{\s v}$ evaluated at the freeze-out temperature.
It is straightforward to conclude that these particles
can achieve the observed DM abundance $\Omega_{\text DM}h^2\simeq 0.12$ \cite{Lahav:2014vza}
as a standard thermal relic for $M\simeq 320\, \GEV$ (assuming, for concreteness, a Dirac fermion, $g=4$).
In such a case, the following hierarchy is necessary: $M_X>T_{\rm max} > T_R > M$ and the
{\it unwanted} relics need to be much heavier $M_X \gtrsim 50\,\TEV$. Thus, as with the SM generations, there is a big gap between the neutral fermion and the charged ones. There is another option: the masses of the new states can have a small hierarchy among them. In this case, the abundances of all
the new particles can be suppressed by lower reheat and maximum temperatures, i.e. by assuming
$M_X> M>T_{\rm max} > T_R $. Now the
DM abundance is given by Eq. (\ref{abundanceNT}). For the same Dirac fermion,
the condition to have $\Omega_{\text DM}h^2\simeq 0.12$ \cite{Lahav:2014vza} reads:
\begin{eqnarray}
\Gamma(9,2x_{\rm max})\,x_R^{-7}\simeq 1.9 \cdot 10^{-19}\,,
\end{eqnarray}
where we have fixed $g_*=106.75$. The lowest hierarchy within the new generation, that allows to have an important
abundance of this type of hypercharged DM, is provided by the exponential suppression with
$x_{\rm max}\simeq x_R$. In this case, $M\simeq 26\,T_R$ is able to account for the total DM, and $M_X\sim 20\, M$
is more than enough to avoid all the anomalous heavy isotope constraints (for $T_R\lesssim 10^6\,\GEV$).

This framework is very interesting, since the new generation can be quite light if one takes into account only cosmological constraints.
In this case, the most promising phenomenology is provided by precision and high energy experiments and it is completely different
with respect to standard DM scenarios. Today, these constraints extend to few hundreds of $\GEV$, but a detailed analysis need
to be performed for this particular case. For collider studies, in contrast to other fourth generation models, the new
quarks and leptons need to be produced by pairs, and in addition, they do not cascade down completely to SM particles. For instance,
events producing the new quarks that give rise to new stable and neutral hadronic states, are expected to contribute with important
amounts of missing energy and transverse momentum. The range of masses that can be excluded at colliders depends in general upon the
path length of the new quark in the detector, the amount of energy deposited in its hadronic collisions, and the probability for
the quark to fragment to an intermediate metastable charged hadron resulting from one of the mentioned hadronic collisions
\cite{Baer:1998pg, Raby:1998xr}.
In the case of a charged stable lepton, constraints can be set directly by analyzing charged tracks
\cite{King:2013kca,PDG}. Standard missing transverse momentum signatures are provided by the production of neutrinos and neutral
leptons \cite{PDG}.
Note that there is also new electrically neutral and charged gauge bosons, that depending on their masses, may be the new stable
states and provide the most distinguishable phenomenology of new {\it right generations}.

Other constraints to stable charged particles come from observations of cosmic rays from satellite or balloon experiments
\cite{Dimopoulos Champs, Barwick, Snowden,Perl Rev Champs}, CMB anisotropies \cite{Dubovsky,Melchiorri, Raffelt Millicharged},
deep underground experiments \cite{Perl Rev Champs}, interstellar clouds diffusion \cite{Chivukula}, and
stellar evolution \cite{Gould}.

New color charged particles change the rate at which the strong interaction
becomes weaker at higher energies: it would become weaker even more quickly.
In a similar way, the new hypercharged states affect the running of the Hypercharge coupling.
These effects can not only constrain the content of new generations, but alternatively, the new matter
content can improve the unification of the couplings at high energies. Note that in the models studied on
this work, the unification should be extended to the new $SU(2)_R$ interaction.

We have already remarked that these models can provide light non-thermal DM candidates. This fact is not
obvious due to the important restrictions which apply to them. However, these DM candidates can also
be very heavy since the abundance given by Eq. (\ref{abundanceNT}) depends fundamentally on the ratio between
their masses and the reheat or maximum temperature. This idea has been already explored in different contexts
\cite{Chung:1998zb,Chung:1998ua,Chung:1998rq,Giudice:2000ex,Feldstein:2013uha}.
Constraints are much weaker, but they may also have associated interesting observational signatures. For instance,
\cite{Feldstein:2013uha} shows that a similar type of heavy DM constitutes an accessible target for direct detection experiments
even for $M\sim 10^{11}\, \GEV$.

It is also important to mention that the stability of the new generation may be broken at the Planck scale
\cite{Krauss:1988zc} or at a lower scale in more complex scenarios by the introduction of new states.
They will effectively introduce non-renormalizable five dimension operator terms, which can also have associated
an interesting phenomenology, for instance, for neutrino physics. In particular, the new neutrinos and even
the neutral lepton may play the role of sterile neutrinos to mix with the SM ones. We consider that all these
novel experimental analyses and theoretical ideas deserve further investigation.

\vspace{0.5cm}
{\bf Acknowledgements}
This work has been supported in part by the MICINN (Spain) project FIS2011-23000, FPA2011-27853-01, Consolider-Ingenio MULTIDARK CSD2009-00064, SNI and CONACYT (Mexico).

\bibliographystyle{ieeetr}

\end{document}